# Anticipations and Discoveries of the Heavy Hydrogen Isotopes, 1913-1939

Helge Kragh*

.

**Abstract:** Just as the chemical elements from hydrogen ($Z = 1$) to oganesson ($Z = 118$) once were discovered, so were the numerous isotopes. The histories of how the isotopes were discovered are less well known, but in a few cases they are as interesting and instructive as those of the elements figuring in the periodic table. Following an overview of criteria usually associated with the concept of discovery in general, this paper examines in detail the historical developments that led to the discoveries of deuterium and tritium and also, as a by-product, the helium-3 isotope. It also includes a brief section on the neutron, which in the 1920s, when it was still a hypothetical particle, was sometimes discussed together with the mass-2 and mass-3 hydrogen isotopes. The paper concludes with a discussion of priority questions relating to suggestions of the two heavy isotopes as well as to their actual discoveries.

## 1. Anticipations, Predictions, and Discoveries

The generally adopted criteria associated with a scientific discovery, whatever its kind, can be summarized in three minimum requirements. A discovery must be (1) novel, (2) presented in public, and (3) supported by a majority of the relevant scientific community [Rancke-Madsen 1975]. None of these criteria are epistemic, which implies, contrary to what has been stated by the philosopher Peter Achinstein [2001], that a claim can be accepted as a discovery at least temporarily even though it is later recognized to be false. Whether concerned with elements, isotopes or something else, the second criterion is a *sine qua non*. A discovery claim must be publicly available, either in the form of a publication or in a well-publicised lecture or the like. To qualify as a discovery, it is not enough that it is known privately or only to a few persons, such as will be exemplified in this essay. It goes without saying that the third criterion presupposes the second one and also, for that matter, that the stated criteria are far from universally accepted.

---

* Niels Bohr Institute, University of Copenhagen, 2100 Copenhagen, Denmark. E-mail: helge.kragh@nbi.ku.dk. This is a revised version of a chapter manuscript which supposedly will appear in a book on the nature of scientific discoveries and in its first version was submitted in 2021.

The notion of discovery is not an either-or category. It also covers the cases where a discovery claim is accepted by a significant fraction of the scientific community but rejected by another and perhaps equally significant fraction. An example of such a 'half-discovery' may be the discovery of triatomic hydrogen $H_3$, to be discussed below. Moreover, both with respect to elements and isotopes there are cases in which a work reported at a certain time is only understood as a proper discovery in retrospect, after it has been complemented or justified by a later work. This kind of dual discovery is related to but not the same as what is usually called a rediscovery. Again, we shall meet one such case below, namely in connection with the discovery of tritium. For a fuller discussion of the discovery concept in relation to chemical elements, see for example [Kragh 2019].

While some discoveries are serendipitous or more or less accidental, others are to some extent derived from theoretical assumptions preceding the actual discovery. The latter are often said to be predicted discoveries, a well-known example being Dmitri Mendeleev's so-called prediction of the elements gallium, scandium, and germanium. However, the term 'prediction' is tricky as it covers a broad range and is often used indiscriminately. Mendeleev's predictions or grounded expectations of new elements are of a different kind than the rigorous theory-derived predictions known from physics and astronomy. Mendeleev's were analogical, not very sharp, and his periodic system did not rely crucially on them. By contrast, Niels Bohr's prediction of the so-called isotope effect to be discussed in the following section was deductively based on a definite theory of atomic structure. Had it turned out that there was no isotope effect, Bohr's atomic theory would have been falsified [Kragh 2012b].

In many cases the term prediction is simply a misnomer which would better be replaced by a more appropriate term such as suggestion, speculation, anticipation, hypothesis, or expectation. Thus, when scientists around 1920 suggested the existence of isotopic particles such as H-2 (deuterium), H-3 (tritium), and He-3, they merely speculated that these entities might exist. The speculation or educated guesswork was not supported by either experiment or theory, the only argument being that the particles were simple proton-electron composites and therefore conceivable entities. To speak of such 'why-not arguments', as we may call them, as predictions is at best an unfortunate terminology.

## 2. Early Suggestions of Mass-3 Isotopes

Given that deuterium (H-2) is stable and relatively abundant, whereas tritium (H-3) is unstable with a half-life of 12.3 years and only exists naturally as trace amounts, one might expect that the possibility of the former isotope was discussed before the latter. But curiously, the history of H-3 precedes that of H-2. Whereas the possible existence of H-3 was suggested as early as September 1913 – a few months before the coining of the word 'isotope' – the possibility of H-2 only turned up in the scientific literature in 1919.

In experiments carried out in Cambridge at about 1912 with positive ions, Joseph John Thomson identified a component with mass 3 which he called $X_3$, and suggested that it was due to an hitherto unknown triatomic hydrogen molecule of the form $H_3$ [Kragh 2012a]. Shortly after having published his groundbreaking paper on atomic theory in the summer of 1913, Bohr attended a meeting of the British Association for the Advancement of Science in Birmingham in which Thomson gave a talk on his $H_3$ hypothesis. In the discussion session that followed the presentation, Bohr suggested as an alternative that $X_3$ might be a mass-3 isotope of hydrogen rather than a triatomic molecule (of course, he did not use the term isotope). The suggestion was not accepted in the scientific community, but it was known to at least some physicists and chemists.

Thus, in a letter to Ernest Rutherford of 14 October 1913, the chemist George Hevesy wrote:

> Bohr – in his usual modest way – suggested the possibility of $X_3$ being an H-atom with one central charge, but having a three times so heavy nucleus than Hydrogen. … Bohr's suggestion is that $X_3$ is possibly a chemically non-separable element from Hydrogen. Of course it is not very probable, but still a very interesting suggestion, which should not be quickly dismissed. [Eve 1939, p. 224].

After having returned to Copenhagen, Bohr realized that the hypothesis might be tested spectroscopically by means of what is known as the 'isotope effect.' According to Bohr's new atomic theory the spectroscopic Rydberg constant given by

$$R_\mathrm{H} = 2\pi^2 \mu e^4 / h^3 c$$

was not a true constant as it depended on the reduced mass $\mu$ of the electron-nucleus system. It followed that the hydrogen spectral lines should depend slightly on the nuclear mass by the factor $(1 + m/M)$, where $m/M$ is the ratio between the mass of the electron and that of the nucleus. He consequently expected the hypothetical H-3 particle, if it existed, to turn up as a very small but measurable shift in wavelength for which he calculated the value

$$\Delta\lambda/\lambda_\text{H} = 2m/(3m + 3M) = 3.6 \times 10^{-4}$$

Bohr looked for this isotope effect, but he found nothing and presumably for this reason he did not refer to the hypothesis in any of his publications. He briefly mentioned the spectroscopic isotope effect in a discussion at the British Association meeting in 1915, but at the time without referring to the heavy hydrogen isotope [Kragh 2012b]. Thus, Bohr's idea of the H-3 isotope was largely unknown to contemporary scientists. The predicted spectral shift for the H-2 isotope would be approximately 1.5 times larger than the one of the H-3 isotope and thus easier to detect, but because Bohr's thoughts were focused on Thomson's $H_3$ molecule, he did not contemplate the possibility of the simpler H-2 isotope.

The H-3 isotope only turned up in the scientific literature in 1920, first in investigations of the isotopic composition of hydrochloric acid (HCl) made by William Harkins, a chemist at the University of Chicago. Harkins [1920a] not only found evidence for the chlorine isotopes of mass 35 and 37 – such as Francis Aston had done slightly earlier – he also searched for what he called the 'meta-hydrogen' of mass 3. Although his search was unsuccessful, he maintained that the corresponding nucleus consisting of three protons and two electrons ($^3_1\text{H}$ or $h_3e^+$ in his notation) was an important unit in the structure of the heavier atomic nuclei. The term 'proton' as a synonym for the hydrogen nucleus or $H^+$ ion was only coined this year and Harkins did not use it [Romer 1997].

In lengthy and rather speculative papers in *Physical Review* and *Journal of the American Chemical Society* Harkins argued that atomic nuclei consisted mainly of hydrogen and helium nuclei, meaning protons and alpha particles. But he also considered other unit particles, including a mass-2 hydrogen nucleus ($^2_1\text{H}$) and two particles of mass 3, one being the $^3_1\text{H}$ particle and the other a nucleus consisting of three protons and one electron, that is, $^3_2\text{He}$. As to the former particle, the nucleus of meta-hydrogen, he wrote that "it would be the nucleus of an atom of a nuclear charge equal to one, and in this sense would be an isotope of hydrogen" [Harkins 1920b, p. 83]. Harkins thought that H-3 might well be real but making up only a miniscule fraction of natural hydrogen. On the other hand, he speculated that it might be represented by the hypothetical element 'nebulium' the existence of which relied on unidentified spectral lines in the distant nebulae and was sometimes ascribed an atomic weight of about 2.7 [Buisson *et al.* 1914]. It took until 1927 before the mysterious nebulium lines were fully explained on the basis of quantum mechanics and then ascribed to $O^{2+}$, an unusual form of the oxygen ion [Hirsh 1979].

Independently of Harkins, in his Bakerian Lecture of 1920 Rutherford reported to have found evidence of the He-3 nucleus in experiments with alpha particles colliding with gaseous molecules such as $O_2$ and $N_2$. The experimental data, he wrote, demonstrated the existence of "atoms of mass 3 carrying two positive charges … [and with] physical and chemical properties very nearly identical with those of helium" [Rutherford 1920, p. 394]. He suggested that the He-3 isotope might be detected by means of the shift in its spectral lines, such as originally proposed by Bohr with respect to H-3.

Not only did Rutherford believe to have found He-3, he also found it "very likely that one electron can also bind two H nuclei and possibly also one H nucleus," which would entail "the possible existence of an atom of mass 2 carrying one charge … an isotope of hydrogen" [Rutherford 1920, p. 396]. Rutherford had considered the possibility of a mass-2 particle already in his early experiments with alpha particles colliding with nitrogen, but without elaborating or commenting on their nature. On 10 January 1918 he entered in his notebook: "Suppose long-range scintillations in $N_2$ are due to atom charge + e and mass M = 2 called *x*" [Feather 1940, p. 152], and in his paper of the following year he cautiously suggested the existence of "atoms of mass 2" [Rutherford 1919, p. 586]. While Rutherford thus anticipated the existence of He-3 and H-2, he did not refer to the possibility of H-3.

The possible existence of two forms of triatomic hydrogen – one a molecule ($H_3$) and the other a heavy isotope (H-3) – led for a while to confusion, not least because Harkins in some of his papers insisted to designate his meta-hydrogen $H_3$. For example, in a report on inorganic chemistry to the Chemical Society the British chemist Edward Baly [1921, p. 34] stated: "Now there seems no doubt that the helium isotope discovered by Rutherford is a different entity from $H_3$, which forms an integral part of Harkins's theory [and] was first discovered by Thomson. … There thus exist two elements of mass 3, one an isotope of hydrogen and the other an isotope of helium." The same confusion was expressed in an address to the British Association given by Edward Thorpe, an elderly chemist and historian of chemistry. According to Thorpe [1921, p. 292], "The hydrogen isotope $H_3$ [meaning H-3], first detected by J. J. Thomson, of which the existence has been confirmed by Aston, would seem to be an integral part of atomic structure … an isotope with an atomic weight of three and nuclear charge of one."

The chief advocates of the $H_3$ molecule, the American chemist Gerald Wendt and his student Robert Landauer, considered the H-3 molecular hypothesis in greater detail than previous authors. In this connection they discussed the possibility of 'iso-hydrogen,' their name for the H-3 isotope. However, they argued that the isotope could not possibly be formed in the chemical reactions responsible for the formation of the $H_3$ molecule [Wendt and Landauer 1922; Kragh 2012a]. As we shall, see, it would take a decade before the H-3 isotope returned on the scene of science.

## 3. Digression: The Neutron as a Chemical Element

Both Rutherford and Harkins suggested around 1920 the existence of a bound proton-electron system, what they called a 'neutron' and which differed in important respects from the later elementary particle of the same name. Rutherford [1920, p. 396] at first conceived the neutron – "an atom of mass 1 which has zero nucleus charge" – as if it were an atom of an unusual chemical element: "Its presence would probably be difficult to detect by the spectroscope, and it may be impossible to contain it in a sealed vessel." Given that the neutron had no external electron it was a strange element indeed, and yet the German chemist Andreas von Antropoff, a professor at the University of Bonn, decided in 1926 to place it in his own version of the periodic table. As a name for this element of atomic number zero he proposed 'neutronium' [Fontana *et al.* 2015, p. 444].

Antropoff's speculation was ignored, but after James Chadwick's discovery of what we now call the neutron in 1932 it was revived by the ever-imaginative Harkins. Chadwick originally believed to have found Rutherford's proton-electron particle and it took more than a year before he realized that he had discovered a new elementary particle with a mass a little larger than the proton-plus-electron system. According to Harkins, Chadwick's neutron was the atom of an extraordinary form of matter, which he called 'neuton' (he also considered the name 'neutronium') and which he believed might be found in the interior of stars. In papers published in *Nature* and *Scientific Monthly*, he wrote enthusiastically about the "element zero, which is totally different from all the other elements … [and] which possesses the unique characteristic of being at the same time an atomic nucleus and a complete atom" [Harkins 1933]. Harkins's speculations of an element zero were not taken more seriously than those of Antropoff. When the astronomers Walter Baade and Fritz Zwicky in 1934 introduced the important idea of neutron stars, which were first observed in the form of pulsars in 1968, they were unaware of Harkins's suggestion.

## 4. The Discovery of Deuterium

The hydrogen isotope of mass 2 was first considered in 1919, if only as a possibility. As mentioned above, in his paper of 1919 Rutherford referred briefly to the idea of a nucleus made up of two protons and one electron, an idea which he had privately entertained the previous year. Slightly later and no less briefly, Harkins [1920b] suggested that atomic nuclei contained units of the same structure, or what in his terminology and nomenclature was 'isohydrogen' (not to be confused with the isohydrogen that Wendt and Landauer considered) with symbol $h_2e$. As yet another constituent of the nucleus he proposed the neutral $h_2e_2$ unit, which in modern physics corresponds to a bound system of two neutrons or what is sometimes referred to as a 'dineutron'. The Rutherford-Harkins $h_2e$ particle corresponds to a deuteron.

Of more interest is a little known contribution by two German scientists, the physicist Otto Stern – a collaborator of Einstein and later a Nobel laureate – and the chemist Max Volmer. In a paper dated 7 January 1919 the two scientists analyzed the deviations of the atomic weights of hydrogen and oxygen from whole numbers, which led them to consider the hypothesis of "an isotopic hydrogen nucleus of atomic weight 2 which possibly may be part of naturally occurring hydrogen" [Stern and Volmer 1919, p. 226]. They calculated that in order to explain the atomic weights, the abundance of the heavy isotope should be approximately 0.8% of natural hydrogen. Moreover, they hypothesized that a mixed hydrogen molecule of molecular weight 3 might be responsible for what Thomson thought was $H_3$. The molecule they had in mind consisted of one ordinary hydrogen atom combined with an atom of the heavy isotope, that is, HD in later terminology. Stern and Volmer reported elaborate diffusion experiments with the purpose of detecting the heavy hydrogen isotope, but without finding any trace of either this isotope or isotopes of oxygen. "We have thus proved," they prematurely concluded, "that … hydrogen and oxygen are not mixtures of isotopes" [Stern and Volmer 1919, p. 238]. In reality, their experiments merely showed that for every million light hydrogen atoms there were less than 10 heavy atoms.

Other researchers in the 1920s agreed with the negative conclusion of Stern and Volmer. Thus, in the second edition of his authoritative monograph *Isotopes*, Aston [1924, p. 72] stated confidently that "it is safe to conclude that hydrogen is a simple element" and that the same was the case with oxygen. Aston's "simple element" was one composed of just a single isotope, a monoisotopic element in later terminology. Only in 1929 did the American chemist William Giauque and his collaborator

Herrick Johnston discover oxygen isotopes with mass numbers 17 and 18 in the atmosphere [Garrett 1962]. It took another two years before the hydrogen mass-2 isotope was taken seriously. Since the discovery of H-2 or deuterium is well described in the literature [Brickwedde 1982; Coffey 2008, pp. 208-221; Clark and Reader 2012], I only recapitulate it in outline.

The Giauque-Johnston discovery of heavy oxygen isotopes indirectly paved the way for the discovery of deuterium [Shindell 2019, p. 75]. In May 1931 the physicist Raymond Birge and the astrophysicist Donald Menzel referred to the possibility of a mass-2 hydrogen isotope, which they thought might explain why the atomic weight of the element came out with the same value on the physical and the chemical mass scale. Whereas the physicists adopted the standard $^{16}O = 16$, the International Commission of Atomic Weights assigned the atomic weight 16 to natural oxygen [Holden 2004]. The recognition that naturally occurring oxygen is polyisotopic, a mixture of $^{16}O$, $^{17}O$ and $^{18}O$, thus implied a small difference between the physical and chemical atomic weight scales.

The idea of Birge and Menzel was essentially the same as the one examined more than a decade earlier by Stern and Volmer, but Birge and Menzel were apparently unaware of the German paper. What matters is that Aston's apparently authoritative value for the hydrogen mass, if expressed in the chemical scale, was 1.00756 and thus lower than the value 1.000776 found by means of chemical precision measurements. Since the discrepancy was outside the limits of error, it posed a problem. It was in this context that Birge and Menzel [1931, p. 1670] pointed out that the discrepancy "could be removed by postulating the existence of an isotope of hydrogen of mass 2, with a relative abundance $H^1/H^2 = 4500$."

Harold Clayton Urey, since 1929 associate professor of chemistry at Columbia University, was an expert in quantum theory and physical chemistry [Shindell 2019]. As evidenced by a massive textbook of 1930 coauthored with Arthur Ruark and titled *Atoms, Molecules and Quanta*, at the time Urey did not presume hydrogen to be composed of more than one isotope. The two authors briefly referred to the old paper by Stern and Volmer, confidently accepting their claim that hydrogen was not a mixture of two or more isotopes. However, the Birge-Menzel paper which appeared the year after caused Urey to change his mind. Immediately after having read the paper, he decided to transform the speculation to a reality by looking systematically for the H-2 isotope. For theoretical reasons he suspected that there existed nuclei with two and three protons bound together by one and two electrons, respectively, or in modern language nuclei of deuterium and tritium. He also suspected a helium isotope of mass 5 (Figure 1).

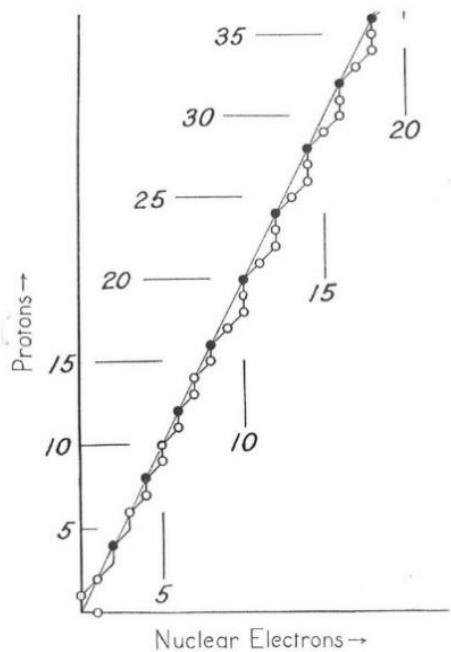

Figure 1. Urey's 1931 chart of protons versus nuclear electrons with filled circles indicating known isotopes and open ones those that would be expected to be found according to the trend shown in the chart. Source: [Urey *et al.*, 1932, p. 14]. Reprinted with permission from the American Physical Society.

Given that the argument of Birge and Menzel turned out to be invalid, as it rested on Aston's wrong value of hydrogen's atomic weight, it is ironic that their incorrect paper served as a direct inspiration for Urey's discovery of deuterium. As Urey [1966, p. 354] stated in a postscript to his Nobel Lecture, without the error "it is probable that … the discovery of deuterium would have been delayed for some time" (see also [Meija 2022]). Looking back on the development, Aston [1935] likewise said: "I am never likely to regret my underestimate of the mass of H made nine years ago, however serious it may ultimately turn out to be, since it played so fundamental a part in encouraging the search for deuterium."

To detect the as yet hypothetical H-2 isotope Urey teamed up with his Columbia colleague George Murphy. Their plan was to look for the characteristic spectral shift in samples of liquid hydrogen prepared by means of fractional distillation by Ferdinand Brickwedde, a physicist at the National Bureau of Standards in Washington D.C. In a careful investigation of the spectral lines Urey and Murphy found shifts in wavelengths which they attributed to a H-2 component. Since the measured shifts agreed with those calculated, they were confident that now they had discovered the mass-2 isotope. For the strongest of the hydrogen lines with wavelength 6563 Å Urey and Murphy observed a spectral shift $\lambda_\text{H} - \lambda_\text{D}$ of 1.79

Å and calculated 1.787 Å on the assumption that it was due to the presence of a mass-2 isotope.

The discovery process concerning the new H-2 discovery was planned, systematic and with no surprises. On the last days of 1931 Urey reported the results in a ten-minute talk at a meeting of the American Physical Society, and the detailed paper of Urey and his collaborators appeared in *Physical Review* on 1 April 1932. According to the three authors, the relative abundance of the new isotope was about 1:4000, of the same order as the presently known ratio 1:6700. Urey, Brickwedde, and Murphy [1932, p. 15] not only looked for H-2 but also for H-3. However, "No evidence for $H^3$ has yet been found, but further concentration … may yet show that this nuclear species exists." The three authors did not discuss the constitution of the H-2 nucleus but implicitly assumed that it was composed by two protons and one electron. After Chadwick's discovery of the neutron later in 1932, the accepted constitution of the deuteron changed from $(2p^+, e^-)$ to $(p^+, n)$.

Whereas the discovery process went smoothly, the naming of the heavy isotope and its nucleus did not. On the contrary, it involved a protracted debate [Stuewer 1986]. A variety of names were proposed and discussed, among them 'diplogen' and 'diplon', which Rutherford preferred, while others argued that the H-2 nucleus should be called a 'deuton', a name frequently used in the period. Still another suggestion was dygen/dyon, such as argued by some American scientists. Urey, on the other hand, was in favour of deuterium/deuteron and for the other two hydrogen isotopes he proposed 'protium' (H-1) and 'tritium' (H-3). Although protium is rarely used, Urey's names eventually won official recognition [IUPAC 1971]. The terms deuterium and deuteron are based on the Greek word 'deuteros' (δεύτερος) for second, whereas 'prote' (πρώτη) means first.

The discovery of deuterium was quickly confirmed by other physicists and chemists, and within a year more than hundred research papers were published on the new isotope and its chemical compounds. Urey's former professor, the prominent Berkeley chemist Gilbert Newton Lewis, was among those who jumped on the bandwagon, quickly writing a series of papers on deuterium and heavy water. According to Urey's biographer, Lewis's sudden activity in deuterium studies was resented by Urey, who felt the competition from Lewis to be unfair [Shindell, 2019, p. 77; see also below].

The importance of the discovery was recognized by the Swedish Academy of Sciences which in 1934 awarded Urey the full Nobel Prize in chemistry "for his discovery of heavy hydrogen." The spectroscopic method that led to the discovery had its origin in Bohr's old theory, although by 1931 this theory had been replaced by the new atomic theory based on quantum mechanics. Actually, the latter theory gives the very same expression for the isotope shift as the older one. In 1923 Urey had studied for a year at Bohr's institute in Copenhagen, but he most likely was unaware of Bohr's early and half-forgotten prediction of the isotope effect. In his Nobel lecture Urey [1966, p. 342], however, mentioned that "Bohr's theory, given some twenty years ago, permits the calculation of the Balmer spectrum of the heavier isotopes of hydrogen by the well-known theoretical formula for the Rydberg constant" (Figure 2).

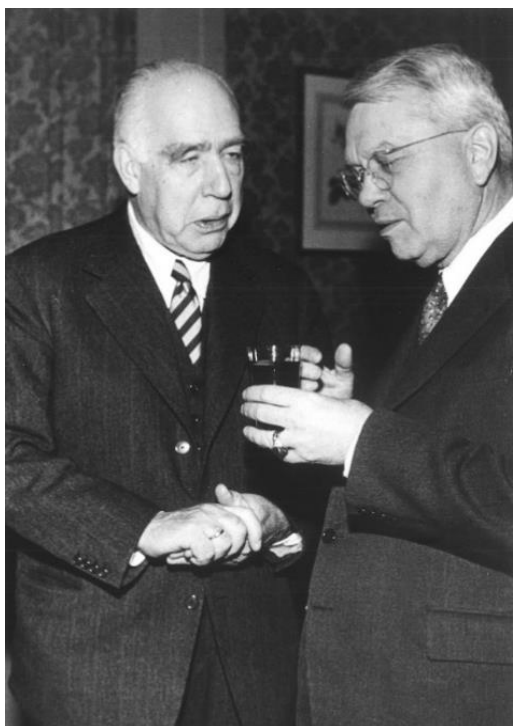

Figure 2. Bohr and Urey in conversation at a meeting in Zurich in 1957.
Credit: Niels Bohr Archive Photo Collection.

Although Urey's work was generally hailed as one of the great discoveries of the twentieth century, not all experts were happy with designating deuterium as an isotope of hydrogen. In a discussion meeting of the Royal Society, none other than Frederick Soddy – who on 4 December 1913 had introduced the notion of isotopy and in 1921 was rewarded with a Nobel Prize for his work – objected that chemical inseparability was a key criterion of isotopy. Without accepting the modern definition of isotopy in terms of nuclear structure he insisted that deuterium was not an isotope of hydrogen but merely a new species of the element. Speaking of "the questionable extension of the now well-understood term 'isotope' to heavy hydrogen," Soddy stated that 'hydrogen isotope' was simply a misnomer [Rutherford *et al*. 1934, p. 11]. As he saw it, since the light and heavy hydrogen could be separated chemically, they were not isotopes of the same element.

## 5. The Roads to Tritium

Although briefly discussed by Harkins and a few other scientists around 1920, the hypothetical H-3 isotope was not taken very seriously at first and seems to have been ignored until the early 1930s, when interest was reawakened by the discovery of deuterium. At about 1930 the American physicist Fred Allison developed a magneto-optical method of chemical analysis by means of which he and his coworkers claimed to have made a number of remarkable discoveries, which included no less than 16 lead isotopes and, most notably, the discovery of elements 85 and 87 which he named "alabamium" and "virginium", respectively [Scerri 2013, pp. 149-152]. Allison [1933, p. 75] also claimed to have found evidence of a heavy hydrogen isotope in 1931, before the work of Urey. However, Allison's method and his use of it was controversial and by the mid-1930s his work was generally disbelieved, eventually to be debunked as pseudoscience [Fontani *et al*. 2015, pp. 327-331]. Indeed, when the American physical chemist Irving Langmuir, a Nobel Prize laureate of 1932, introduced the concept of 'pathological science' as another name for pseudoscience, Allison's work featured as a prime example [Langmuir 1989].

Before Allison was debunked, the Berkeley chemist Wendell Latimer made magneto-optical experiments with water solutions containing about 3% deuterium, reporting that he had found evidence of H-3 [Latimer and Young 1933]. However, the claim was contradicted by careful spectrographic experiments with 67% concentrated heavy water made by G. N. Lewis and his assistant Frank Spedding. Failing to find any trace of the expected spectral shift within the sensitivity of their apparatus, the two chemists concluded that the H-3 isotope "appears not to be present in ordinary hydrogen to the extent of 1 part in 6,000,000" [Lewis and Spedding 1933, p. 964]. Nor did painstaking studies with the recently invented mass spectrometer reveal any sign of the superheavy hydrogen isotope.

If H-3 were absent from water and also from the solar spectrum (where Lewis and Spedding had also looked for it), perhaps it might turn up in nuclear reactions? British and American physicists thought so and in the spring of 1934 the plot thickened with several claims of having detected the elusive particle. By that time the atomic nucleus no longer consisted of protons and electrons, but of protons and neutrons. At the Cavendish Laboratory, Rutherford and his assistants, the Australian physicist Mark Oliphant and the Austrian physical chemist Paul Harteck, bombarded deuterium compounds with deuterons accelerated in the laboratory's new high-voltage generator. The experiments were principally due to Oliphant and Harteck, while Rutherford's contribution was largely limited to discussion and interpretation of the results (Figure 2). The three researchers suggested as one possibility the reaction

$$^2_1H + {^2_1}H \rightarrow {^4_2}He^* \rightarrow {^1_1}H + {^3_1}H$$

with the release of 4.0 MeV of energy. The second possibility was that the unstable $^4He^*$ nucleus decayed into a He-3 nucleus and a neutron according to

$$^2_1H + {^2_1}H \rightarrow {^4_2}He^* \rightarrow {^3_2}He + {^1_0}n$$

In this case the surplus energy would be 3.2 MeV. Not only did Rutherford and his two coworkers believe they had detected the new isotopes H-3 and He-3, they also thought they had found a stable Be-8 isotope in some of the nuclear reactions. This isotope is in fact exceedingly unstable, with a lifetime of only $10^{-16}$ s. Nonetheless, some twenty years later it came to play an important role in nuclear astrophysics, namely as an intermediary nuclide in the synthesis of carbon-12 from three alpha particles, $3\,{^4_2}He \rightarrow {^{12}_6}C$.

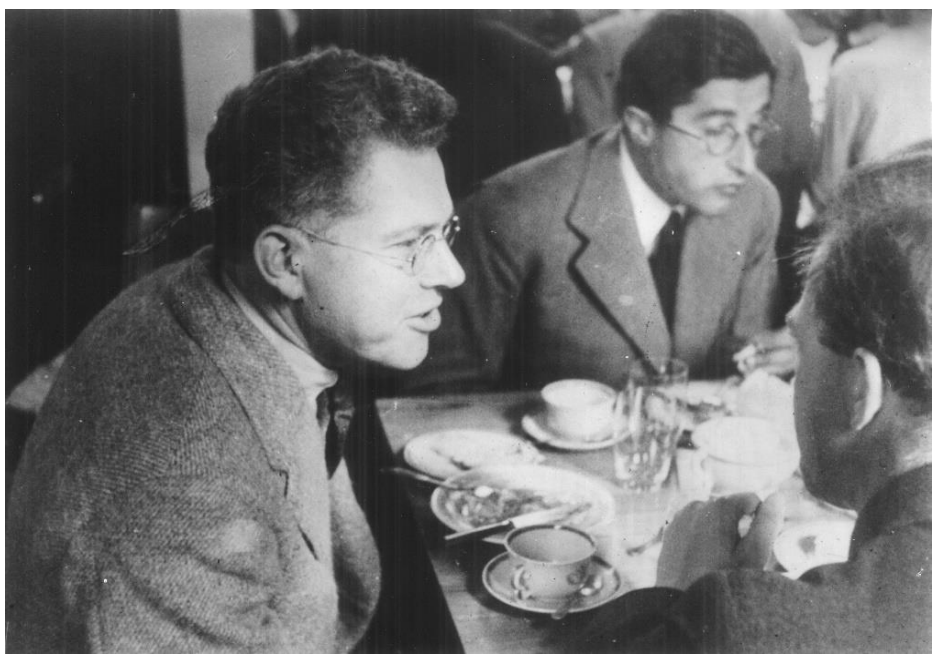

Figure 3. Mark Oliphant, co-discoverer of tritium, at the cafeteria of Bohr's institute during a conference in 1936. The person to his right is Czech physicist George Placzek. Credit: Niels Bohr Archive Photo Collection.

The Cavendish group published a preliminary report in *Nature* of 17 March 1934 in which they cautiously stated that the two mass-3 isotopes "appear to be stable for the short time required for their detection, [but] the question of their permanence requires further consideration" [Oliphant *et al*. 1934a]. Rutherford judged the potential discoveries of the two isotopes in somewhat different terms. As he confided in a letter to Ernest Lawrence of 13 March 1934, "I personally believe that there can be little doubt of the reaction in which the hydrogen isotope of mass 3 is produced, … [while] the evidence for the helium isotope of mass 3 is … at present somewhat uncertain" [Dahl 2002, p. 155]. In their detailed paper in the *Proceedings of the Royal Society* dated a month later, Rutherford and his collaborators similarly expressed confidence in the discovery of H-3, whereas their evaluation of the mass-3 helium isotope was more circumspect: "While we have not yet detected the $_2$He$^3$ particles which we believe to be present, we have not yet obtained any evidence that they do not exist" [Oliphant *et al*. 1934b, p. 702].

American physicists were competing with and closely following the work done at the Cavendish Laboratory. They might have been beaten in the race to find the H-3 isotope, but what about its presence in nature? Analyzing a sample of nearly pure deuterium with an advanced mass spectrometer, in April 1934 a group of three Princeton physicists reported evidence for a mass-5 component, which they interpreted as due to a binary DT molecule consisting of one H-2 atom (D) and one H-3 atom (T). Without using the name 'tritium' they designated H-3 with the symbol T. The estimated abundance was exceedingly low, namely about one H-3 atom per one billion of H-1 atoms. Nonetheless, the result "confirms rather satisfactorily the existence of a third isotope of hydrogen from natural sources" [Lozier *et al*. 1934].

Shortly later another team of American physicists reported evidence for H-3 atoms, this time based on the study of nuclear reactions roughly of the same kind as those made at the Cavendish Laboratory. Merle Tuve and his coworkers at the Carnegie Institution at Washington D.C. used high-speed deuterons to bombard a highly concentrated deuterium target sample, and by measuring the range of the resulting particles they found evidence for H-3 atoms. Whereas Rutherford's group concluded that the H-3 nuclei were the result of a nuclear transformation, Tuve's group suggested that they were stable constituents present in the H-2 target and consequently that their work provided "satisfactory proof of the existence of H$^3$ atoms in deuterium samples" [Tuve *et al*. 1934, p. 841]. Unfortunately, nobody else could reproduce the results communicated by Tuve and his group.

The situation in the fall of 1934 was summarized in a popular paper with the title "Protium-Deuterium-Tritium: The Hydrogen Trio" written by Hugh Taylor, a chemistry professor at Princeton University. According to Taylor [1934, p. 367] there was no doubt that tritium – "the youngest and rarest of all the isotopes" – was part of the natural world. But even so, tritium compounds would never become commodities of any practical use: "Pure tritium water, $T_2O$, would cost at least $10,000,000 a gram or [tritium] water roughly twenty times the cost of radium."

In a paper published in *Nature* on 21 August 1937, which sadly turned out to be his last one, Rutherford carefully appraised the history and current status of H-3 and He-3, referring to H-3 as 'triterium' rather than tritium. He maintained the Cavendish priority to the discoveries of the two isotopes, but now concluded that they rapidly disappeared in experiments and that none of them existed naturally. Yet he did not suspect any of the isotopes to be radioactive. On the instigation of Rutherford, Aston had used his mass spectrograph to look for the H-3 isotope in a sample of pure heavy water supplied by the Norwegian company Norsk Hydro. However, contrary to what Rutherford expected, Aston's careful analysis showed no sign of the mass-3 isotope. Rutherford [1937, p. 305] commented: "It does not seem feasible at the moment to obtain sufficient quantities of these two interesting isotopes to study properties by ordinary physical chemical methods." This was not the end of the discovery history, though, for it can be argued that tritium was not discovered in 1934 but only five years later.

In 1939 the young Berkeley physicist Luis Alvarez, who 29 years later would be awarded the Nobel Prize for his contributions to elementary particle physics, and his graduate student Robert Cornog used the Berkeley cyclotron to produce a steady beam of 24-Mev He-3 ions. They demonstrated that traces of the rare isotope were present in ordinary helium and estimated an abundance ratio relative to atmospheric helium of about $10^{-7}$. Alvarez and Cornog [1939] (see also [Cornog, 1987]) further showed that the gas mixture produced in the deuteron-deuterium process was radioactive, which they ascribed to a H-3 component decaying into He-3. The beta process in later nomenclature is

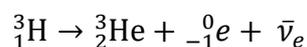

$$^{3}_{1}\text{H} \rightarrow {}^{3}_{2}\text{He} + {}^{0}_{-1}e + \bar{\nu}_e$$

On the other hand, they found the He-3 nucleus to be stable. Alvarez and Cornog also investigated the half-life of tritium in their 1939 experiments, but without coming up with a definite result. The two physicists merely estimated a lifetime "greater than ten years" [Alvarez and Cornog 1940], whereas the first determination of tritium's half-life following in 1940 gave a value of 31 ± 8 years. The modern value is 12.32 ± 0.02 years. Deuterium as well as tritium proved immensely important, but the later consequences of the discoveries related to thermonuclear weapons are outside the scope of the present paper [Rhodes 1995].

## 6. Summary and Discussion

The hydrogen gas or 'inflammable air' discovered by Henry Cavendish in 1766 is a mixture of two isotopes, such as was demonstrated in 1932 with the discovery of deuterium which makes up about 0.031% by mass of naturally occurring hydrogen. A few years later the two hydrogen isotopes were extended with a third one, the radioactive tritium of mass 3. Although tritium exists in nature, its abundance in hydrogen is close to zero, namely about $10^{-18}$ % and its presence mainly due to nuclear reactions in the upper atmosphere caused by cosmic rays. Here, a fast neutron may react with nuclei of atmospheric nitrogen according to

$$^{14}_{7}\text{N} + ^{1}_{0}n \rightarrow ^{12}_{6}\text{C} + ^{3}_{1}\text{H}$$

Deuterium (D) and tritium (T) are the only ones among the numerous isotopes of different elements which are still granted their own names and chemical symbols. As stated by IUPAC, the International Union of Pure and Applied Chemistry, in 1957: "All isotopes of an element should have the same name. For hydrogen the isotope names protium, deuterium and tritium, may be retained, but it is undesirable to assign isotopic names instead of numbers to other elements" [Bassett 1960; Connolly and Damhus, 2005]. IUPAC's decision to disallow isotope names except those of hydrogen, indicates the special significance of the two light isotopes.

The historical actors involved in the discovery processes were chemists as well as physicists, some being physical chemists and others nuclear physicists. Generally speaking, much of the work was interdisciplinary. However, as far as the discovery of tritium is concerned, it largely took place in laboratories devoted to nuclear physics and with little or no participation of traditionally laboratory-trained chemists.

Whereas the discoveries of new chemical elements are officially recognized either by IUPAC or informally by the chemical community, this is not generally the case with the numerous isotopes belonging to the elements. Just as astronomers do not celebrate the discovery of yet another comet, or yet another exoplanet, so chemists and physicists consider the discovery of yet another isotope to be relatively unimportant. There are too many of them to make headlines in the history of science. While the discovery of a new element has in several cases been recognized with a Nobel Prize, witness radium, polonium, plutonium and the inert gases, only a single isotope discovery has been similarly rewarded, namely Urey's discovery of deuterium. (Giauque's 1949 Nobel Prize was for his work in low-temperature physics and not for the discoveries of the oxygen isotopes.) The discovery of deuterium was fairly unproblematic and has never been seriously questioned nor confronted with rival discovery claims. Urey and his collaborators Murphy and Brickwedde did not refer to their work on deuterium as a 'discovery' in any of their papers of 1932, but the Nobel Committee had little doubt that it qualified as one.

Although Urey's paternity to deuterium was broadly accepted, his full Nobel Prize for the discovery was far from self-evident. There was another serious candidate for the prize, namely G. N. Lewis whose important work on heavy water and deuterium chemistry was highly considered by the Nobel Committee in Stockholm. In a report to the committee of 18 May 1934 Theodor Svedberg, a reputed Swedish physical chemist and himself a Nobel laureate of 1926, stated that "the main merit for the discovery is without doubt ascribed to Urey" [Coffey 2008, p. 217]. Nonetheless, Svedberg ended up with recommending the prize to be shared by Urey and Lewis. Only after further deliberations did he change his mind, now arguing that Lewis's work was after all derivative to that of Urey. To the disappointment of Lewis, when the prize was announced on 15 November 1934 it went to Urey alone [Coffey 2008, p. 221; Shindell 2019, p. 78]. It could have been divided between him and his two collaborators Murphy and Brickwedde, but this possibility was not contemplated in Stockholm. Apparently Urey thought they deserved credit for their contributions to the discovery, for he generously gave half of his Nobel Prize money to be divided between them [Coffey 2008, p. 2019; Shindell 2019, p. 81].

The *idea* of a heavy hydrogen isotope of mass 2 was not new, as it was discussed at a few occasions around 1920 after which it largely disappeared. Rutherford and Harkins independently hypothesized the existence of a mass-2 nuclear particle, but their suggestions were speculative and unconnected to potential experiments. The only pre-discovery experiments aimed at the H-2 isotope were made in Berlin by Stern and Volmer, who concluded that the isotope did not exist. In agreement with what has been argued in the first part of this chapter, none of these early cases can reasonably be characterized as predictions of the discovery made in 1932. They were merely loose anticipations of a speculative nature.

As far as tritium is concerned, Bohr suggested in 1913 that the triatomic hydrogen molecule proposed by Thomson in his positive-ray experiments might be an H-3 isotope. In an interview of 1 November 1962, shortly before his death, Bohr recalled his role in the prehistory of tritium:

> I just took up the question of whether in hydrogen one could have what you now call tritium. And then I saw that it was a way to show this by its diffusion in palladium. Hydrogen and tritium will behave similarly but the masses are so different that they will be separated out. [Kragh 2012b, p. 178].

Although the suggestion was scarcely noticed at the time, it was made in public and counts as a proper prediction. On the other hand, Bohr's slightly later and more important idea of detecting H-3 by spectroscopic means was not known to the scientific community and is therefore irrelevant in a social discovery context. And yet it is of interest retrospectively, given that this 'private discovery claim' might under other circumstances have been communicated to the scientists and thus constituted a proper discovery claim. Insofar that Harkins in 1920 presented his idea of 'meta-hydrogen' as a mass-3 isotope and actually looked for it, he may be considered as having anticipated tritium.

The case of tritium is significantly different from the one of deuterium, not only because of the scarcity of this unstable and more exotic isotope but also because there are two possible candidates for having discovered tritium. According to *Encyclopaedia Britannica* and several other sources, "Tritium was discovered in 1934 by the physicists Ernest Rutherford, M. L. Oliphant, and Paul Harteck" [https://www.britannica.com/science/tritium]. However, there are also sources which credit Alvarez and Cornog with the discovery. The 1939 paper by Alvarez and Cornog did not present their work as a discovery of a new isotope, and yet, in his recollections many years later, Cornog [1987, p. 26] stated that he and Alvarez "discovered hydrogen and helium of mass 3."

Was tritium discovered twice? There are indeed cases in which elements or something else can reasonably be said to have been discovered twice or even thrice (such as the discovery of chlorine or aluminium), but in my view tritium does not belong to this category. As shown by the historical record, the Cavendish group demonstrated the existence of the H-3 nucleus in their experiments of 1934 and, what is no less important, the claim was generally accepted by the relevant scientific communities. It is worth noting that strictly speaking the Cavendish group did not discover the atomic species tritium but only its nucleus sometimes known as a triton, a word (τριτον) meaning third in Greek. In this respect there is a parallel to the much later artificially produced 'superheavy elements' of atomic number $Z > 103$. These have in some cases been recognized as real, and hence being discovered, on the basis of the detection of only a few atomic nuclei [Kragh 2018]. There is also an interesting analogy to the discovery of technetium ($Z = 43$), which in 1937 was synthesized by Emilio Segré and Carlo Perrier as the first ever element produced in the laboratory [Scerri 2013, pp. 116-143]. It is often forgotten that tritium was similarly produced three years earlier. The analogy goes further: as tritium is not totally absent in nature, so is it the case with technetium.

As far as the work of Alvarez and Cornog is concerned, the two physicists undoubtedly made a discovery, but not of a new hydrogen isotope. They discovered that tritium is radioactive and also that the He-3 isotope is part of natural helium, but with regard to tritium they merely confirmed its existence. It is hard to avoid the conclusion that priority to the discovery of tritium belongs to Rutherford and his group at the Cavendish Laboratory. Nonetheless, it was only with the Alvarez-Cornog work that tritium became fully recognized as an important object in nuclear physics. The discovery histories of the heavy hydrogen isotopes illustrate the problematic relationship between the related but yet different concepts of anticipation, prediction and discovery. In the cases of both deuterium and tritium there were earlier suggestions and theoretically based expectations of the possible existence of the isotopes. The first group may qualify as anticipations, the second as predictions, but whatever their names they were not discoveries. In a nutshell, they missed the crucial element of experimental detection widely accepted by the scientific community as the hallmark of true discovery.